\input{aipcheck}

\documentclass[
    ,final            % use final for the camera ready runs
  ]
  {aipproc}

\layoutstyle{6x9}

\begin{document}

\title{Average and recommended half-life values for two neutrino double beta decay: upgrade-09}


\author{A.S. Barabash}{
  address={Institute of Theoretical and Experimental Physics, B.\
Cheremushkinskaya 25, 117259 Moscow, Russia}
}

\begin{abstract}
 All existing ``positive'' results on two neutrino double beta decay in 
different nuclei were analyzed.  Using the procedure recommended by the 
Particle Data Group, weighted average values for half-lives of 
$^{48}$Ca, $^{76}$Ge, $^{82}$Se, $^{96}$Zr, $^{100}$Mo, $^{100}$Mo - 
$^{100}$Ru ($0^+_1$), $^{116}$Cd, $^{130}$Te, $^{150}$Nd, $^{150}$Nd - $^{150}$Sm 
($0^+_1$) and $^{238}$U were obtained. Existing geochemical data were 
analyzed and recommended values for half-lives of $^{128}$Te, $^{130}$Te 
and $^{130}$Ba are proposed.  We recommend the use of these results as
presently the most precise and reliable values for half-lives.
\end{abstract}

\maketitle

%%%%%%%%%%%%%%%%%%%%%%%%%%%%%%%%%%%%%%%%%%%%
%% MAINMATTER
%%%%%%%%%%%%%%%%%%%%%%%%%%%%%%%%%%%%%%%%%%%%

\section{Introduction}

  At present, the two neutrino double beta ($2\nu\beta\beta$) decay process 
has been detected in a total of 10 different nuclei. In $^{100}$Mo and 
$^{150}$Nd, this type of decay was also detected for the transition to
the $0^+$ excited state of the daughter nucleus. For the case of the 
$^{130}$Ba nucleus, evidence for the two neutrino double electron capture 
process was observed via a geochemical experiment.  All of these results 
were obtained in a few tens of geochemical experiments, more then thirty 
direct (counting) experiments, and in one radiochemical experiment. In 
direct experiments, for some nuclei there are as many as seven independent 
positive results (e.g., $^{100}$Mo).  In some experiments, the statistical 
error does not always play the primary role in overall half-life 
uncertainties. For example, the NEMO-3 experiment with $^{100}$Mo detected 
more than 219000 useful events \cite{ARN05}, which results in a value for 
the statistical error of $\sim$ 0.2\% . At the same time, the systematic 
error in many other experiments on $2\nu\beta\beta$ decay generally 
remains quite high ($\sim 10-30\%$) and very often cannot be determined 
very reliably.  As a result, it is frequently quite difficult for the 
``user'' to select the ``best'' half-life value among all existing 
results.  In fact, however, using an averaging procedure, one can produce 
reliable and accurate half-life values for each isotope.

In the present work, a critical analysis of all ``positive'' experimental
results has been performed, and averaged (or recommended) values for all 
isotopes have been obtained.

The first time that this type of work was done was in 2001, and the 
results were presented at MEDEX'01 \cite{BAR02}. Then upgrated half-life values were
presented at MEDEX'05 \cite{BAR06}. In the present paper, 
new positive results obtained since 2005 have been added and analyzed.  
\begin{table}
\caption{Present,``positive'' $2\nu\beta\beta$ decay results. 
Here, N is the number of useful events, S/B is the signal-to-background 
ratio. $^{*)}$ For $E_{2e} > 1.2$ MeV. $^{**)}$ After correction (see text). 
$^{***)}$ For SSD mechanism. $^{****)}$ In both peaks.}
\bigskip
\label{Table1}
%%\begin{tabular*}{\textwidth}{l@{\extracolsep{\fill}}cccc@{   }ccc}
%\begin{tabular*}{\textwidth}{l|cccc|c@{\extracolsep{\fill}}cc}
\begin{tabular}{|c|c|c|c|c|}
\hline
\rule[-2.5mm]{0mm}{6.5mm}
Nucleus & N & $T_{1/2}$, y & S/B & Ref., year \\
%\rule[-1mm]{0mm}{3.5mm} 
\hline
\rule[-2mm]{0mm}{6mm}
$^{48}$Ca & $\sim 100$ & $[4.3^{+2.4}_{-1.1}(stat)\pm 1.4(syst)]\cdot 10^{19}$
  & 1/5 & \cite{BAL96}, 1996 \\
 & 5 & $4.2^{+3.3}_{-1.3}\cdot 10^{19}$ & 5/0 & \cite{BRU00}, 2000 \\
& 116 & $[4.4^{+0.5}_{-0.4}(stat)\pm 0.4(syst)\cdot 10^{19}$ & 6.8 & \cite{FLA08}, 2008 \\
\rule[-4mm]{0mm}{10mm}
 & & {\bf Average value:} $\bf 4.4^{+0.6}_{-0.5} \cdot 10^{19}$ & & \\  
          
\hline
\rule[-2mm]{0mm}{6mm}
$^{76}$Ge & $\sim 4000$ & $(0.9\pm 0.1)\cdot 10^{21}$ & $\sim 1/8$                                                        
& \cite{VAS90}, 1990 \\
& 758 & $1.1^{+0.6}_{-0.3}\cdot 10^{21}$ & $\sim 1/6$ & \cite{MIL91}, 1991 \\
& 132 & $0.93^{+0.2}_{-0.1}\cdot 10^{21}$ & $\sim 4$ & \cite{AVI91}, 1991 \\
& 132 & $1.2^{+0.2}_{-0.1}\cdot 10^{21}$ & $\sim 4$ & \cite{AVI94}, 1994 \\
& $\sim 3000$ & $(1.45\pm 0.15)\cdot 10^{21}$ & $\sim 1.5$ & \cite{MOR99}, 1999 
\\
& $\sim 80000$ & $[1.74\pm 0.01(stat)^{+0.18}_{-0.16}(syst)]\cdot 10^{21}$ & $\sim 1.5$ 
& \cite{HM03}, 2003 \\
\rule[-4mm]{0mm}{10mm}
& & {\bf Average value:} $\bf (1.5\pm 0.1) \cdot 10^{21}$ & & \\

\hline
\rule[-2mm]{0mm}{6mm}
$^{82}$Se & 149.1 & $[0.83 \pm 0.10(stat) \pm 0.07(syst)]\cdot 10^{20}$ & 2.3 & 
\cite{ARN98}, 1998 \\
& 89.6 & $1.08^{+0.26}_{-0.06}\cdot 10^{20}$ & $\sim 8$ & \cite{ELL92}, 1992 \\
& 2750 & $[0.96 \pm 0.03(stat) \pm 0.1(syst)]\cdot 10^{20}$ & 4 & \cite{ARN05}, 2005\\ 
& & $(1.3\pm 0.05)\cdot 10^{20}$ (geochem.) & & \cite{KIR86}, 1986 \\
\rule[-4mm]{0mm}{10mm}
& & {\bf Average value:} $\bf (0.92\pm 0.07)\cdot 10^{20}$ & & \\
 
\hline
\rule[-2mm]{0mm}{6mm}
$^{96}$Zr & 26.7 & $[2.1^{+0.8}_{-0.4}(stat) \pm 0.2(syst)]\cdot 10^{19}$ & $1.9^{*)}$ 
& \cite{ARN99}, 1999 \\
& 453 & $[2.35 \pm 0.14(stat) \pm 0.19(syst)]\cdot 10^{19}$ & 1 & \cite{FLA08}, 2009\\
& & $(3.9\pm 0.9)\cdot 10^{19}$ (geochem.)& & \cite{KAW93}, 1993 \\
& & $(0.94\pm 0.32)\cdot 10^{19}$ (geochem.)& & \cite{WIE01}, 2001 \\
\rule[-4mm]{0mm}{10mm}
& & {\bf Average value:} $\bf (2.3 \pm 0.2)\cdot 10^{19}$ & & \\

\hline
\rule[-2mm]{0mm}{6mm}
$^{100}$Mo & $\sim 500$ & $11.5^{+3.0}_{-2.0}\cdot 10^{18}$ & 1/7 & 
\cite{EJI91}, 1991 \\
& 67 & $11.6^{+3.4}_{-0.8}\cdot 10^{18}$ & 7 & \cite{ELL91}, 1991 \\
& 1433 & $[7.3 \pm 0.35(stat) \pm 0.8(syst)]\cdot 10^{18**)}$ & 3 & 
\cite{DAS95}, 1995 \\
& 175 & $7.6^{+2.2}_{-1.4}\cdot 10^{18}$ & 1/2 & \cite{ALS97}, 1997 \\
& 377 & $[6.75^{+0.37}_{-0.42}(stat) \pm 0.68(syst)]\cdot 10^{18}$ & 10 & 
\cite{DES97}, 1997 \\
& 800 & $[7.2 \pm 1.1(stat) \pm 1.8(syst)]\cdot 10^{18}$ & 1/9 & 
\cite{ASH01}, 2001 \\
& 219000 & $[7.11 \pm 0.02(stat) \pm 0.54(syst)]\cdot 10^{18***)}$ & 40 & 
\cite{ARN05}, 2005\\
& & $(2.1\pm 0.3)\cdot 10^{18}$ (geochem.)& & \cite{HID04}, 2004 \\ 
\rule[-4mm]{0mm}{10mm}
& & {\bf Average value:} $\bf (7.1\pm 0.4)\cdot 10^{18}$ & & \\

\hline
\end{tabular}
\end{table}

%Table 1 continued
\addtocounter{table}{-1}
\begin{table}
\caption{continued.}
\bigskip
%\label{Table1}
\begin{tabular}{|c|c|c|c|c|}

\hline
%\rule[-2mm]{0mm}{6mm}
$^{100}$Mo - & $133^{****)}$ & $6.1^{+1.8}_{-1.1}\cdot 10^{20}$ & 1/7 & 
\cite{BAR95}, 1995 \\
$^{100}$Ru ($0^+_1$) &  $153^{****)}$ & $[9.3^{+2.8}_{-1.7}(stat) \pm 1.4(syst)]\cdot 
10^{20}$ & 1/4 & \cite{BAR99}, 1999 \\
 & 19.5 & $[5.9^{+1.7}_{-1.1}(stat) \pm 0.6(syst)]\cdot 10^{20}$ & $\sim 8$ & 
\cite{DEB01}, 2009 \\ 
& 35.5 & $[5.5^{+1.2}_{-0.8}(stat) \pm 0.3(syst)]\cdot 10^{20}$ & $\sim 8$ & 
\cite{KID09}, 2009 \\ 
& 37.5 & $[5.7^{+1.3}_{-0.9}(stat) \pm 0.8(syst)]\cdot 10^{20}$ & $\sim 3$ & 
\cite{ARN07}, 2007 \\   
\rule[-4mm]{0mm}{10mm}
& & {\bf Average value:} $\bf 5.9^{+0.8}_{-0.7}\cdot 10^{20}$ & & \\

\hline
%\rule[-2mm]{0mm}{6mm}
$^{116}$Cd& $\sim 180$ & $2.6^{+0.9}_{-0.5}\cdot 10^{19}$ & $\sim 1/4$ & 
\cite{EJI95}, 1995 \\
& 9850 & $[2.9\pm 0.06(stat)^{+0.4}_{-0.3}(syst)]\cdot 10^{19}$ & $\sim 3$ & 
\cite{DAN03}, 2003 \\
& 174.6 & $[2.9 \pm 0.3(stat) \pm 0.2(syst)]\cdot 10^{19**)}$ & 3 & 
\cite{ARN96}, 1996 \\
& 1370 & $[2.8 \pm 0.1(stat) \pm 0.3(syst)]\cdot 10^{19***)}$ & 7.5 & \cite{FLA08}, 2008\\
\rule[-4mm]{0mm}{10mm}
& & {\bf Average value:} $\bf (2.8 \pm 0.2)\cdot 10^{19}$ & & \\

\hline
\rule[-2mm]{0mm}{6mm}
$^{128}$Te& & $\sim 2.2\cdot 10^{24}$ (geochem.) & & \cite{MAN91}, 1991 \\
& & $(7.7\pm 0.4)\cdot 10^{24}$ (geochem.)& & \cite{BER93}, 1993 \\
& & $(2.41\pm 0.39)\cdot 10^{24}$ (geochem.)& & \cite{MES08}, 2008 \\
& & $(2.3\pm 0.3)\cdot 10^{24}$ (geochem.)& & \cite{THO08}, 2008 \\
\rule[-4mm]{0mm}{10mm}
& & {\bf Recommended value:} $\bf (2.0\pm 0.3)\cdot 10^{24}$ & & \\

\hline
\rule[-2mm]{0mm}{6mm}
$^{130}$Te& 260 & $[6.1 \pm 1.4(stat)^{+2.9}_{-3.5}(syst)]\cdot 10^{20}$ & 1/8 & \cite{ARN03}, 2003 \\
& 236 & $[6.9 \pm 0.9(stat)^{+1.0}_{-0.7}(syst)]\cdot 10^{20}$ & 1/3 & \cite{TRE09}, 2009 \\
& & $\sim 8\cdot 10^{20}$ (geochem.) & & \cite{MAN91}, 1991 \\
& & $(27\pm 1)\cdot 10^{20}$ (geochem.)& & \cite{BER93}, 1993 \\
& & $(9.0\pm 1.4)\cdot 10^{20}$ (geochem.)& & \cite{MES08}, 2008 \\
& & $(8.0\pm 1.1)\cdot 10^{20}$ (geochem.)& & \cite{THO08}, 2008 \\
\rule[-4mm]{0mm}{10mm}
& & {\bf Recommended value:} $\bf (6.8\pm 1.2)\cdot 10^{20}$ & & \\

\hline
\rule[-2mm]{0mm}{6mm}
$^{150}$Nd& 23 & $[18.8^{+6.9}_{-3.9}(stat) \pm 1.9(syst)]\cdot 10^{18}$ & 
1.8 & \cite{ART95}, 1995 \\
& 414 & $[6.75^{+0.37}_{-0.42}(stat) \pm 0.68(syst)]\cdot 10^{18}$ & 6 & 
\cite{DES97}, 1997 \\
& 2018 & $[9.11^{+0.25}_{-0.22}(stat) \pm 0.63(syst)]\cdot 10^{18}$ & 2.8 & \cite{ARG08}, 2008\\
\rule[-4mm]{0mm}{10mm}
& & {\bf Average value:} $\bf(8.2\pm 0.9)\cdot 10^{18}$ & & \\

\hline
\rule[-2mm]{0mm}{6mm}
$^{150}$Nd - & $177.5^{****)}$ & $[1.33^{+0.36}_{-0.23}(stat)^{+0.27}_{-0.13}(syst)]\cdot 10^{20}$ & 
1/5 & \cite{BAR09}, 2009 \\
$^{150}$Sm ($0^+_1$) & & {\bf Average value:} $\bf 1.33^{+0.45}_{-0.26}\cdot 10^{20}$ & \\ 
 
\hline
\rule[-2mm]{0mm}{6mm}
$^{238}$U& & $\bf (2.0 \pm 0.6)\cdot 10^{21}$ (radiochem.) & & \cite{TUR91}, 1991 \\
 
\hline
\rule[-2mm]{0mm}{6mm}
$^{130}$Ba &  & $\bf (2.2 \pm 0.5)\cdot 10^{21}$ (geochem.) & 
 & \cite{MES01}, 2001 \\
ECEC(2$\nu$) & & & \\ 

\hline
\end{tabular}
\end{table}

\section{Present experimental data}

Experimental results on $2\nu\beta\beta$ decay in different nuclei are 
presented in Table 1.  For direct experiments, the number of useful events 
and the signal-to-background ratio are presented.

\section{Data analysis}

To obtain an average of the available data, a standard weighted 
least-squares procedure, as recommended by the Particle Data Group 
\cite{PDG00}, was used.  The weighted average and the corresponding error 
were calculated, as follows:
\begin{equation}
\bar x\pm \delta \bar x = \sum w_ix_i/\sum w_i \pm (\sum w_i)^{-1/2} , 
\end{equation} 
where $w_i = 1/(\delta x_i)^2$.  Here, $x_i$ and $\delta x_i$ are, 
respectively, the value and error reported by the i-th experiment, and 
the summations run over the N experiments.  

The following step is to calculate $\chi^2 = \sum w_i(\bar x - x_i)^2$ and 
compare it with N - 1, which is the expectation value of $\chi^2$ if the 
measurements are from a Gaussian distribution.  If $\chi^2/(N - 1)$ is 
less than or equal to 1, and there are no known problems with the data, 
we accept the results.  If $\chi^2/(N - 1)$ is very large, we may choose 
not to use the average at all.  Alternatively, we may quote the calculated 
average, while making an educated guess of the error, using a conservative 
estimate designed to take into account known problems with the data.
Finally, if $\chi^2/(N - 1)$ is larger than 1 but not greatly so, we may 
still average the data, but can increase the quoted error, $\delta \bar x$ 
in Equation 1, by a scale factor S defined as 
\begin{equation}
S = [\chi^2/(N - 1)]^{1/2}.
\end{equation} 
For averages, we add the statistical and systematic errors in quadrature 
and use this combined error as $\delta x_i$. In some cases only the results 
obtained with high enough 
signal-to-background ratio were used.

\subsection{$^{48}$Ca }    
There are three independent experiments in 
which $2\nu\beta\beta$ decay of $^{48}$Ca was observed \cite{BAL96,BRU00,FLA08}. 
The results are in good agreement. The weighted average value is:
$$
T_{1/2} = 4.4^{+0.6}_{-0.5} \cdot 10^{19} y.
$$ 

\subsection{$^{76}$Ge } 
Let us consider the results of five 
experiments. First of all, however, a few additional comments are 
necessary:

1) Result of the Heidelberg-Moscow group was corrected. 
Instead of the previously published value $T_{1/2} = [1.55\pm 
0.01(stat)^{+0.19}_{-0.15}(syst)]\cdot 10^{21}$ y \cite{KLA01}, a new 
value $T_{1/2} = [1.74\pm 0.01(stat)^{+0.18}_{-0.16}(syst)]\cdot 10^{21}$ y
 \cite{HM03} has been presented. It is the latter value that has been used 
in our present analysis.  At the same time, using an independent analysis, 
the Moscow part of the Collaboration obtained a value similar to the result 
of Ref. \cite{HM03}, namely $T_{1/2} = 
[1.78\pm 0.01(stat)^{+0.08}_{-0.10}(syst)]\cdot 10^{21}$ y \cite{BAK03}.

2) In Ref. \cite{AVI91}, the value $T_{1/2} = 
0.92^{+0.07}_{-0.04}\cdot 10^{21}$ y was presented. However, after a more 
careful analysis, this result has been changed to a value of 
$T_{1/2} = 1.2^{+0.2}_{-0.1}\cdot 10^{21}$ y \cite{AVI94}, 
which was used in our analysis.

3) The results presented in Ref. \cite{VAS90} do not agree with the more 
recent and more precise experiments \cite{HM03,MOR99}.   Furthermore, the 
error presented in \cite{VAS90} appears to be too small, especially taking 
into account the fact that the signal-to-background ratio in this 
experiment is equal to $\sim 1/10$. It has been mentioned before
\cite{BAR90} that the half-life value in this work can be $\sim 1.5-2$ 
times higher because the thickness of the dead layer in the Ge(Li) 
detectors used can be different for crystals made from enriched Ge, rather 
than natural Ge. With no uniformity of the external background, this 
effect can have an appreciable influence on the final result.

Finally, in calculating the average, only the results of experiments 
with signal-to-background ratios greater than 1 were used (i.e., the 
results of Refs. \cite{HM03,AVI94,MOR99}). The weighted average value is:
$$
    T_{1/2} = (1.5 \pm 0.1) \cdot 10^{21} y.
$$ 

\subsection{$^{82}$Se}
There are three independent counting 
experiments and many geochemical measurements $(\sim 20)$. The geochemical 
data are neither in good agreement with each other nor in good agreement 
with the data from direct measurements.  Formally, the accuracy of 
geochemical measurements is typically on the level of a few percent and
sometimes even better.  Nevertheless, the possibility of existing large 
systematic errors cannot be excluded (see discussion in Ref. \cite{MAN86}). 
It is mentioned in Ref. \cite{BAR00} that if the weak interaction constant 
$G_F$ is time-dependent, then the half-life values obtained in geochemical 
experiments will depend on the age of the samples.  Thus, to obtain a 
``present'' half-life value for $^{82}$Se, only the results of the direct 
measurements \cite{ARN05,ARN98,ELL92} were used.  The result of Ref. 
\cite{ELL87} is the preliminary result of \cite{ELL92}, hence it has not
been used in our analysis.  It is interesting to note that the ``lower'' 
error in Ref. \cite{ELL92} appears to be too small.  Indeed, it 
is even smaller than the statistical error, and that is why we use here a 
more realistic value of 15\%  as an estimation of this error. As a result, 
the weighted average value is:
$$
T_{1/2} = (0.92 \pm 0.07) \cdot 10^{20} y.
$$ 

\subsection{$^{96}$Zr} 
There are two ``positive'' geochemical results
\cite{KAW93,WIE01} and two results from direct NEMO-2 \cite{ARN99} and 
NEMO-3 \cite{FLA08} experiments.  Taking into account the comment in 
section 3.3, we use the values from Refs. \cite{ARN99,FLA08} to obtain 
a ``present'' weighted half-life value for $^{96}$Zr of: 
$$
T_{1/2} = (2.3 \pm 0.2)\cdot 10^{19} y.                    
$$ 

\subsection{$^{100}$Mo} 
Formally, there are seven positive 
results\footnote{We do not consider the result of Ref. \cite {VAS90a} 
because a possible high background contribution to the ``effect'' was 
not excluded in this experiment.} from direct experiments and one recent
result from a geochemical experiment. However, we do not consider the 
preliminary result of M. Moe et al. \cite{ELL91} and instead use their 
final result \cite{DES97}, plus we do not use the geochemical result 
(again, see comment in section 3.3).  Finally, in calculating the average, 
only the results of experiments with signal-to-background
ratios greater than 1 were used (i.e., the results of Refs. 
\cite{DAS95,DES97,ARN05}).  In addition, here we have used the corrected 
half-life value from Ref. \cite {DAS95}.  First of all, the original 
result was decreased by 15\% because the calculated efficiency (by MC)
was overestimated (see Ref. \cite {VAR97}).  Secondly, the half-life 
value was decreased by 10\% taking into account that, for the special 
case of $^{100}$Mo we have to deal with the Single State Dominance (SSD) 
mechanism (see discussion in \cite {ARN04,SHI06}). The following weighted average value 
for this half-life is obtained:
$$
T_{1/2} = (7.1 \pm 0.4)\cdot 10^{18} y .                                   
$$
In framework of High State Dominance (HSD) mechanism (see \cite{SIM01,DOM05}) the following 
average value can be obtained, $T_{1/2} = (7.6 \pm 0.4)\cdot 10^{18}$ y . 

\subsection{$^{100}$Mo - $^{100}$Ru ($0^+_1$; 1130.29 keV)} 
The 
transition to the $0^+$ excited state of $^{100}$Ru was detected in five 
independent experiments.  The results are in good agreement, and the 
weighted average value for half-life is (using results from \cite{BAR95,BAR99,KID09,ARN07}):
$$
T_{1/2} = 5.9^{+0.8}_{-0.7} \cdot 10^{20} y .
$$                                   
The result from \cite{DEB01} was not used here because we consider the result from \cite{KID09}
as final result of TUNL-ITEP experiment.

\subsection{$^{116}$Cd} 
There are three independent ``positive'' 
results that are in good agreement with each other when taking into 
account the corresponding error bars.  Again, we use here the corrected 
result for the half-life value from Ref. \cite{ARN96}.  The original 
half-life value was decreased by $\sim$ 25\% (see remark in section 3.5). The 
weighted average value is (SSD mechanism): 
$$          
T_{1/2} = (2.8 \pm 0.2)\cdot 10^{19} y.
$$ 
If the HSD mechanism is realised for the case of $^{116}$Cd as well, then
the adjusted half-life value is $T_{1/2} = (3.0 \pm 0.2)\cdot 10^{19}$ y. 

\subsection{$^{128}$Te and $^{130}$Te} 
Long time there were only geochemical 
data for these isotopes. Although the half-life 
ratio for these isotopes has been obtained with good accuracy $(\sim 3\%)$ 
\cite{BER93}, the absolute values for $T_{1/2}$ of the individual nuclei 
are different from one experiment to the next.  One group of authors 
\cite{MAN91,TAK66,TAK96} gives $T_{1/2} \approx 0.8\cdot 10^{21}$ y  
for $^{130}$Te and $T_{1/2} \approx  2\cdot 10^{24}$ y for $^{128}$Te, 
whereas another group \cite{KIR86,BER93} claims $T_{1/2} \approx 
(2.5-2.7)\cdot 10^{21}$ y and  $T_{1/2} \approx 7.7\cdot 10^{24}$ y, 
respectively. Furthermore, as a rule, experiments with ``young'' 
samples ($\sim 100$ million years) result in half-life values of 
$^{130}$Te in the range of $\sim (0.7-0.9)\cdot 10^{21}$ y,
while for ``old'' samples ($> 1$ billion years), half-life values in the
range of $\sim (2.5-2.7)\cdot 10^{21}$ y have been produced. 
It was even assumed that the difference in half-life values could be 
connected with a variation of the weak interaction constant $G_F$ with 
time \cite{BAR00}.

One can estimate the absolute half-life values for $^{130}$Te 
and $^{128}$Te using only very well-known ratios from geochemical 
measurements and the ``present'' half-life value of $^{82}$Se (see 
section 3.3).  The first ratio is 
given by $T_{1/2}(^{130}{\rm Te})/T_{1/2}(^{128}{\rm Te}) = 
(3.52 \pm 0.11)\cdot 10^{-4}$ \cite{BER93}, while the second ratio is 
given by $T_{1/2}(^{130}{\rm Te})/T_{1/2}(^{82}{\rm Se}) = 9.9 \pm 0.6$. 
This latter value is the weighted average value from three experiments 
with minerals containing both elements (Te and Se): $7.3 \pm 0.9$ 
\cite{LIN86}, $12.5 \pm 0.9$ \cite{KIR86} and $10 \pm 2$ \cite{SRI73}. 
It is significant that the gas retention age problem has no effect on 
the half-life ratios.  Now, using the ``present'' $^{82}$Se half-life 
value $T_{1/2} = (0.92 \pm 0.07)\cdot 10^{20}$ y and the value $9.9 \pm 
0.6$ for the $T_{1/2}(^{130}{\rm Te})/T_{1/2}(^{82}{\rm Se})$ ratio, one 
can obtain the half-life value for $^{130}$Te:
$$          
T_{1/2} = (9 \pm 1)\cdot 10^{20} y.
$$ 

Using $T_{1/2}(^{130}{\rm Te})/T_{1/2}(^{128}{\rm Te}) = 
(3.52 \pm 0.11)\cdot 10^{-4}$ 
\cite{BER93}, one can obtain the half-life value for $^{128}$Te:
$$          
T_{1/2} = (2.5 \pm 0.3)\cdot 10^{24} y.
$$ 
Recently it was argued that "short"half-lives are more likely to be correct \cite{MES08,THO08}.
Using different "young" mineral results half-life value was estimated as 
$(9.0 \pm 1.4)\cdot 10^{20}$ y \cite{MES08} and $(8.0 \pm 1.1)\cdot 10^{20}$ y \cite{THO08} 
for $^{130}$Te and $(2.41 \pm 0.39)\cdot 10^{24}$ y \cite{MES08} and $(2.3 \pm 0.3)\cdot 10^{24}$ y \cite{THO08} 
for $^{128}$Te (corresponding to the observed $T_{1/2}(^{130}{\rm Te})/T_{1/2}(^{128}{\rm Te})$
ratio).

The first indication of a positive result for $^{130}$Te in a direct experiment was obtained in 
\cite{ARN03}. More accurate and reliable value was obtaind recently in NEMO-3 experiment \cite{TRE09}.
The results are in good agreement, and the waighted average value for half-life is
$$
T_{1/2} = (6.8 \pm 1.2)\cdot 10^{20} y.
$$ 
And now, using the $T_{1/2}(^{130}{\rm Te})/T_{1/2}(^{128}{\rm Te})$
ratio, one can obtane half-life value for $^{128}$Te,
$$
T_{1/2} = (2.0 \pm 0.3)\cdot 10^{24} y.
$$  
We recommend to use these last two results as most precise and reliable half-life values for 
$^{130}$Te and $^{128}$Te, respectively.

\subsection{$^{150}$Nd}
The half-life value was measured in three 
independent experiments \cite{ART95,DES97,ARG08}. Using Equation 1, and three 
existing values one 
can obtain $T_{1/2} = (8.2 \pm 0.5)\cdot 10^{18}$ y.  Taking into account 
the fact that $\chi^2 > 1$ and S = 1.89 (see Equation 2) we finally obtain:
$$
T_{1/2} = (8.2 \pm 0.9)\cdot 10^{18} y.
$$ 

\subsection{$^{150}$Nd - $^{150}$Sm ($0^+_1$; 740.4 keV)}
There is 
only one positive result from a direct (counting) experiment \cite{BAR09}:
$$          
T_{1/2} = [1.33^{+0.36}_{-0.23}(stat)^{+0.27}_{-0.13}(syst)]\cdot 10^{20} y.
$$ 
Preliminary result of this work was published in \cite{BAR04}.
\subsection{$^{238}$U}  
There is only one positive result from 
a radiochemical experiment \cite{TUR91}:
$$          
T_{1/2} = (2.0 \pm 0.6)\cdot 10^{21} y.
$$ 

\subsection{$^{130}$Ba (ECEC)}
There is only one positive result from 
a geochemical experiment \cite{MES01}:
$$          
T_{1/2} = (2.2 \pm 0.5)\cdot 10^{21} y.
$$

\section{Conclusion}

In summary, all ``positive'' $2\nu\beta\beta$-decay results were analyzed 
and average values for half-lives were calculated. For the cases of 
$^{128}$Te and $^{130}$Te, so-called ``recommended'' values 
have been proposed.  We strongly recommend the use of these values as 
presently the most precise and reliable. In particular, the accurate 
experimental $2\nu\beta\beta$-decay rates can be used to adjust the most 
relevant parameter in the framework of the QRPA model, namely the strength 
of the particle-particle interaction ($g_{pp}$). Once accomplished, these 
values can be used in NME calculations for neutrinoless double beta decay 
\cite{ROD06,KOR07}.

%%%%%%%%%%%%%%%%%%%%%%%%%%%%%%%%%%%%%%%%%%%%
%% Sample figure:
%%
%% The option [height=...] scales the picture to the given height,
%% without it it would be printed at its nominal size
%%%%%%%%%%%%%%%%%%%%%%%%%%%%%%%%%%%%%%%%%%%%

%%%%%%%%%%%%%%%%%%%%%%%%%%%%%%%%%%%%%%%%%%%%
%% SAMPLE TABLE
%%
%% Shows the use of \tablehead and \tablenote
%% macros
%%%%%%%%%%%%%%%%%%%%%%%%%%%%%%%%%%%%%%%%%%%%

%%%%%%%%%%%%%%%%%%%%%%%%%%%%%%%%%%%%%%%%%%%%%%%%
%% BACKMATTER
%%%%%%%%%%%%%%%%%%%%%%%%%%%%%%%%%%%%%%%%%%%%%%%%



\bibliographystyle{aipproc}   % if natbib is available

\bibliography{sample}

\IfFileExists{\jobname.bbl}{}
 {\typeout{}
  \typeout{******************************************}
  \typeout{** Please run "bibtex \jobname" to optain}
  \typeout{** the bibliography and then re-run LaTeX}
  \typeout{** twice to fix the references!}
  \typeout{******************************************}
  \typeout{}
 }

\end{document}